\documentclass[journal, twocolumn, twoside, 10pt]{IEEEtran}
\usepackage{amssymb,graphicx,psfrag,cite}
\usepackage{amsmath}
\usepackage{epsfig}
\usepackage{srcltx}
\usepackage{color}
\usepackage{stfloats}
\usepackage{mathrsfs}

\newtheorem{proposition}{Proposition}

\newtheorem{lemma}{Lemma}

\begin{document}


\title{High Throughput Random Access via Codes on Graphs: Coded Slotted ALOHA}
\pagestyle{empty}

\author{Enrico Paolini, Gianluigi Liva, and Marco Chiani
\thanks{E. Paolini and M. Chiani are with DEIS/WiLAB, University of Bologna, via Venezia 52, 47521
Cesena (FC), Cesena, Italy (e-mail: e.paolini@unibo.it, marco.chiani@unibo.it).}%
\thanks{Gianluigi Liva is with German Aerospace Center (DLR), D-82234 Wessling, Germany (e-mail: 
Gianluigi.Liva@dlr.de).}
\thanks{This work was supported in part by the EC under Seventh FP grant agreement ICT
OPTIMIX n.INFSO-ICT-214625.}
}

\maketitle

\thispagestyle{empty}

\begin{abstract}
In this paper, coded slotted ALOHA (CSA) is introduced as a powerful random access scheme to the
MAC frame. In CSA, the burst a generic user wishes to transmit in the MAC frame is first split into
segments, and these segments are then encoded through a local a packet-oriented code prior to
transmission. On the receiver side, iterative interference cancellation combined with decoding of
the local code is performed to recover from collisions. The new scheme generalizes the previously
proposed irregular repetition slotted ALOHA (IRSA) technique, based on a simple repetition of the
users' bursts. An interpretation of the CSA interference cancellation process as an iterative
erasure decoding process over a sparse bipartite graph is identified, and the corresponding density
evolution equations derived. Based on these equations, asymptotically optimal CSA
schemes are designed for several rates and their performance for a finite number of users
investigated through simulation and compared to IRSA competitors. Throughputs as high as $0.8$ are
demonstrated. The new scheme turns out to be a
good candidate in contexts where power efficiency is required.
\end{abstract}

\section{Introduction}
Although demand assignment multiple access (DAMA) medium access control (MAC) protocols guarantee an
efficient usage of the available bandwidth
\cite{abramson94:multiple}, MAC random access schemes remain an appealing and popular solution
for wireless networks. Among them, slotted ALOHA (SA)
\cite{Abramson:ALOHA,Roberts72:ALOHA,abramson94:multiple}
is currently adopted as the initial access scheme in both cellular terrestrial and satellite
communication networks \cite{Morlet07_RCS}. In \cite{Rappaport83:DSA} an improvement to SA was
proposed, namely, diversity slotted ALOHA (DSA). In DSA, each packet (also called \emph{burst})
is transmitted twice over the MAC frame, which provides a slight throughput gain over SA. As a
drawback, for the same peak transmission power of the SA scheme, the average transmitted power
of DSA is doubled. 

A more effective use of the burst repetition is provided by contention resolution diversity slotted
ALOHA (CRDSA) \cite{DeGaudenzi07:CRDSA},
whose basic idea is the adoption of interference cancellation (IC) to resolve collisions. More
specifically, with respect
to DSA, each of the twin replicas of a burst, transmitted within a MAC frame, possesses a
pointer to the slot position where the respective copy was sent. Whenever a clean burst is detected
and successfully decoded, the pointer is extracted and the interference contribution caused by the
burst copy on the corresponding slot is removed. This procedure is iterated, possibly allowing to
recover the whole set of bursts transmitted within the same MAC frame. This results in a
remarkably improved normalized throughput $S$ (defined as the probability of successful packet
transmission per time slot) which may reach $S\simeq0.55$, whereas the peak throughput for pure
SA is $S=1/e\simeq 0.37$. Further improvements can be achieved by exploiting the capture effect
\cite{Roberts72:ALOHA,DeGaudenzi09:CRDSA}.

In \cite{Liva_CRDSA10_SCC,Liva2010:IRSA} irregular repetition
slotted ALOHA (IRSA) was introduced to provide a further
throughput gain over CRDSA. A higher normalized throughput is achieved by IRSA by allowing
a variable and judiciously designed repetition rate for each burst. As for DSA, the performance
improvement achieved by CRDSA/IRSA has a counterpart in the increment of the average
transmitted power. Since CRDSA is a specific instance of IRSA, in the following we will
refer in general to IRSA. In \cite{Liva_CRDSA10_SCC} it is also illustrated how the iterative
burst recovery process on the receiver side can be represented via a bipartite graph and how,
under the assumption of an ideal channel estimation and of a sufficiently large signal-to-noise
ratio (SNR), it shares
several commonalities with the graph representation of the erasure recovery process of modern
channel codes on sparse graphs \cite{luby01:efficient,studio3:Richardson_design_of_cap_appr}. 

In this paper, we introduce a further generalization of IRSA, named \emph{coded slotted
ALOHA} (CSA). The basic idea of CSA is to encode (instead of simply repeat) bursts using
local codes prior to transmission in the MAC frame and to combine, on the receiver side,
iterative IC with decoding of the local codes to recover from collisions. The new scheme turns
out to be interesting especially in contexts where power efficiency is required. Density evolution
equations for CSA are derived to analyze the IC process in an asymptotic setting, leading
to the calculation of the peak asymptotic throughput. Numerical results are then presented to
illustrate the validity of the proposed asymptotic analysis and its effectiveness in the design of
CSA access schemes for a finite number of users.

\section{System Model}\label{sec:system_model}
Similarly to \cite{DeGaudenzi07:CRDSA,Liva_CRDSA10_SCC}, we consider a random access scheme where
the slots are grouped in MAC frames, all with the same length (in slots). We further restrict
to the case where each user attempts one burst transmission per MAC frame. 

Consider $M$ users, each attempting the transmission of a burst of time duration $T_{\mathrm{SA}}$
over a MAC frame of time duration $T_{\mathrm{F}}$. Neglecting guard times, the MAC frame
is composed of $N_{\mathrm{SA}}=T_{\mathrm{F}}/T_{\mathrm{SA}}$ slots. In SA, each user would
independently choose one of the $N_{\mathrm{SA}}$ slots uniformly at random and would attempt
transmission of his burst into that slot. In IRSA each user would generate a certain number $r$
of replicas of his burst, where $r$ may be not the same for two different users, and would transmit
the $r$ replicas into $r$ slots chosen uniformly at random among the available $N_{\mathrm{SA}}$
slots. 

In CSA, when a user wishes to transmit a burst of time duration $T_{\mathrm{SA}}$ over the
MAC frame, the burst is divided into $k$ information sub-bursts (also called information
\emph{segments}), each of time duration $T_{\mathrm{CSA}}=T_{\mathrm{SA}}/k$. The $k$ information
segments are then encoded by the user via a packet-oriented binary linear block code which generates
$n_h$ encoded segments, each of time duration $T_{\mathrm{CSA}}=T_{\mathrm{SA}}/k$. For each
transmission, the code to be employed is drawn randomly by the user from a set of $n_c$ possible
codes. For $h\in\{1,\dots,n_c\}$ the $h$th code, denoted by $\mathscr{C}_h$, is a 
$(n_h,k,d_{\min}^{(h)})$ code, that is it has length $n_h$, dimension $k$, and minimum distance
$d_{\min}^{(h)}$. We further impose that $\mathscr{C}_h$ has no idle bits and fulfills
$d_{\min}^{(h)}\geq2$. We assume that, at any transmission, each user independently chooses his
local code according to a probability mass
function (p.m.f.) {\boldmath $P$}$=[P_h]_{h=1}^{n_c}$ which is the same for all
users. Denoting again by $T_{\mathrm{F}}$ the MAC frame duration, the MAC frame is
composed of $N_{\mathrm{CSA}}=T_{\mathrm{F}}/T_{\mathrm{CSA}}=kN_{\mathrm{SA}}$ slots. The $n_h$
coded segments are then transmitted by the user over $n_h$ slots picked uniformly at random. Note
that IRSA may be seen as a special case of CSA where $k=1$ and each $\mathscr{C}_h$ is a
repetition code of length $n_h$, and that SA is a special case of IRSA where $n_h=1$ for
all users.\footnote{We point out that CSA may be seen as a generalization also of the schemes
proposed in \cite{massey78:ciss,Lam1990:THFHMA}, where no IC was used.} The overall \emph{rate}
of CSA is defined as $R=k/\bar{n}$, where $\bar{n} := \sum_{h=1}^{n_c}P_h n_h$ is the
expected length of the code employed by the generic user. Note that $\Delta P=\bar{n}/k=1/R$
represents the increment of average power with respect to pure~SA.

\begin{figure}[!t]
\psfrag{label1}[c]{\footnotesize{$N_{\rm CSA}$ sum nodes}} \psfrag{label2}[c]{\footnotesize{$M$
burst nodes}}
\begin{center}
\includegraphics[width=0.7\columnwidth,draft=false]{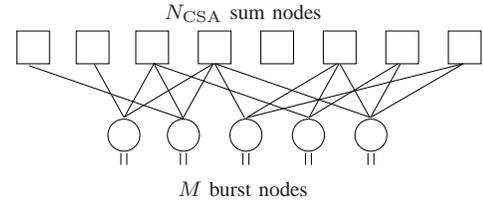}
\end{center}
\caption{Graph representation of CSA. Circles are the burst nodes and represent the $M$ users,
squares are the sum nodes and represent the $N$ slots. The degree of a burst node is equal to the
length of the locally employed code. The degree of a sum node is equal to the number of collided
encoded segments. The example is for $k=2$.}\label{fig:CSA_graph}
\end{figure}

It is now convenient to introduce a graph representation of CSA, depicted in
Fig.~\ref{fig:CSA_graph}. Let us consider a MAC frame composed of $N_{\rm CSA}$
slots, in which $M$ users attempt a transmission. The MAC frame status can be represented by a
bipartite graph, $\mathcal{G} = (\mathcal{B},\mathcal{S},\mathcal{E})$, consisting of a set
$\mathcal{B}$ of $M$ \emph{burst nodes} (one for each burst transmitted in the MAC
frame), a set $\mathcal{S}$ of $N_{\rm CSA}$ \emph{sum nodes} (one for each slot), and a set
$\mathcal{E}$ of edges. An edge connects a burst node (BN) $b_i \in \mathcal{B}$ to a sum node
(SN) $s_j \in \mathcal{S}$ if and only if an encoded segment associated with the $i$th burst is
transmitted in the $j$th slot. In other words, BNs correspond to bursts, SNs to slots,
and edges to encoded segments. Therefore, a burst split into $k$ information segments and encoded
via the code $\mathscr{C}_h$ is represented as a BN with $n_h$ neighbors. Correspondingly, a
slot where $d$ replicas collide is represented as a SN with $d$ connections. The number of
edges emanating from a node is the node degree. Moreover, a BN where $\mathscr{C}_h$ is
employed during the current transmission is referred to as a BN of type~$h$.

In our analysis, we rely on three assumptions. 1) \emph{Sufficiently high SNR}. This allows to
claim that, when a segment is received in a clean slot, it is known at the receiver. 2) \emph{Ideal
channel estimation}. Under this assumption (and the previous one), ideal IC is
possible, allowing the recovery of collided bursts with a probability that is essentially one. 3)
\emph{Destructive collisions}. Segments that collide in a slot are treated as erasures. These
assumptions simplify the analysis without substantially affecting the obtained results, as shown
in \cite{DeGaudenzi07:CRDSA} and \cite{Liva2010:IRSA} for CRDSA and IRSA, respectively.

Each coded segment associated with a BN of type $h$ is equipped with information about the
relevant user and with a pointer to the other $n_h-1$ segments.\footnote{In practical
implementations, the overhead due to
the inclusion of pointers in the segment header may be reduced by adopting more efficient
techniques. For fixed $k$, one may include in the segment header the code index $h$ together with a
random seed, out of which it is possible to reconstruct (by a pre-defined pseudo-random number
generator) the positions of the $n_h$ segments.} On the receiver side, segments which collided in
some slot with those sent by another user are marked as lost, so that a BN is
connected to ``known'' edges and to ``unknown'' ones. Hence, some of its information
bursts are known, and the others unknown. At the generic BN (say of type $h$), erasure
decoding of the code $\mathscr{C}_h$ may allow to recover some of the unknown encoded and
information segments. It is now possible to subtract the interference contribution of the newly
recovered encoded segments from the signal received in the corresponding slot. If $d-1$ segments
that collided in a SN of degree $d$ have been recovered by the corresponding BNs, the
remaining segment becomes known. The IC process combined with local decoding at the BNs
proceeds iteratively, i.e., cleaned segments may allow solving other collisions. Note that
this procedure is equivalent to iterative decoding of a doubly-generalized low-density
parity-check (D-GLDPC) code over the erasure channel \cite{paolini10:DGLDPC_random}, where variable
nodes are generic linear block codes and check nodes are single parity-check (SPC) codes. 

Denoting by $N=N_{\textrm{CSA}}$ the number of slots (a multiple of $k$), the logical normalized
offered traffic $G$ is given by\footnote{In CSA and IRSA we distinguish the
\emph{logical} load $G$ from the \emph{physical} load given by $\frac{\bar{n}}{k}G=G/R$ and
representing the average number of transmitted segments per slot. The logical load $G$ provides a
direct measure of the traffic handled by the scheme. Note that the two concepts coincide in pure
SA.}
\begin{align}\label{eq:offered_traffic}
G=\frac{kM}{N}\, .
\end{align}
The normalized throughput $S$ is defined as the probability of successful packet transmission
per time slot. For example, for standard SA we have $S=Ge^{-G}$.

Finally, we recall the definition of \emph{information function} of a linear block code
\cite{helleseth97:information}. Let $\mathbf{G}$ be a generator matrix for an $(n,k)$
linear block code $\mathscr{C}$. The $g$th un-normalized information function of $\mathscr{C}$,
denoted by $\tilde{e}_g$, is defined as the summation of the ranks over all the possible submatrices
obtained selecting $g$ columns (with $0 \leq g \leq n$) out of $\mathbf{G}$.

\section{Density Evolution, Threshold, and Stability}\label{sec:density_evolution}

The degree distribution of the SNs from a node perspective is defined as
\begin{align}\label{eq:Psi}
\Psi(x)=\sum_{d \geq 0} \Psi_d x^d
\end{align}
where $\Psi_d$ is the probability that a SN has degree $d$. 

Let us consider a user encoding his segments through the code $\mathscr{C}_h$, and allocating his
$n_h$ encoded segments into $n_h$ slots chosen randomly. Then, the probability that the BN
associated with this user (say $U$) is connected to a SN $A$  may be expressed as the
ratio between the number of ways of connecting the $n_h$ sockets of $U$ to the $N$ SNs such
that $U$ is connected to $A$, to the total number of ways of connecting the $n_h$ sockets of $U$ to
the $N$ SNs: $\Pr\{U\textrm{ is connected to }A | U\textrm{ uses }\mathscr{C}_h\}  = {N-1
\choose n_h-1}/{N \choose n_h}= \frac{n_h}{N}$. Therefore we have:
\begin{align*}
\Pr\{U\textrm{ is connected to }A\} & = \sum_{h=1}^{n_c} P_h \frac{n_h}{N}= \frac{\bar{n}}{N}\,.
\end{align*}

Since each user selects his slots independently of all the other users, the probability $\Psi_d$
that a SN has degree $d$ (that is the probability that the SN is chosen by $d$ users) is
given by
\begin{align}
\Psi_d & = {M \choose d} \left(\frac{\bar{n}}{N}\right)^d \left(1-\frac{\bar{n}}{N}\right)^{M-d} \notag \\
\, & \stackrel{\textrm{(a)}}{=} {M \choose d} \left(\frac{\bar{n}G}{kM}\right)^d \left(1-\frac{\bar{n}G}{kM}\right)^{M-d} \notag \\
\, & \rightarrow \frac{e^{-\frac{\bar{n}}{k}G}}{d!} \left(\frac{\bar{n}}{k}\,G\right)^d \quad \textrm{as } M\rightarrow\infty
\end{align}
where equality (a) follows from \eqref{eq:offered_traffic}. Therefore, in the limit where $M$ (and
consequently, for fixed $G$ and $k$, $N$ through \eqref{eq:offered_traffic}) tends to infinity,
\eqref{eq:Psi} may be
written~as
\begin{align}\label{eq:Psi_infty}
\Psi(x) & = \sum_{d\geq 0} \frac{e^{-\frac{\bar{n}}{k}G}}{d!} \left(\frac{\bar{n}}{k}\,G\right)^d
x^d = \exp\left(-\frac{\bar{n}}{k}G(1-x)\right) \, .
\end{align}
Using \eqref{eq:Psi_infty} we can now express the probability $\rho_d$ that an edge is connected to
a SN of degree $d \geq 1$ as:
\begin{align}
\rho_d & = \frac{\Psi_d\, d}{\sum_{i\geq 1}\Psi_i\, i} = \frac{\Psi_d\, d}{\Psi'(1)} =
\frac{\left(\frac{\bar{n}}{k}\,G\right)^{d-1}}{(d-1)!}\,e^{-\frac{\bar{n}}{k}\,G} \, .
\end{align}
Therefore, the degree distribution of the SNs from an edge perspective is given by
\begin{align}\label{eq:rho(x)}
\rho(x) & = e^{-\frac{\bar{n}}{k}\,G} \sum_{d\geq 1}
\frac{\left(\frac{\bar{n}}{k}\,Gx\right)^{d-1}}{(d-1)!} = \exp\left(-\frac{\bar{n}}{k}G(1-x)\right) 
\end{align}
and $\rho(x)=\Psi(x)$.

For given $k$ and $G$, we investigate the evolution of the decoding process
described in Section~\ref{sec:system_model} in the asymptotic case where $M \rightarrow \infty$ (and
consequently $N \rightarrow \infty$ through~\eqref{eq:offered_traffic}).
\medskip
\begin{proposition}\label{prop:qhi}
Assume MAP decoding is used at each BN. At the $i$th decoding iteration, let $p_{i-1}$ be
the average probability that an edge carries an erasure message\footnote{This is the probability
that an edge is associated with an encoded segment that is still unknown.} from the SNs to the
BNs. Consider a BN where $\mathscr{C}_h$ is employed and let $q^{(h)}_i$ be the average
probability that an edge carries an erasure message outgoing from the BN, after MAP decoding
at the BN. Then, we have
\begin{align}\label{eq:q^h_i(p_i-1)}
q^{(h)}_i = & \,\frac{1}{n_h}\sum_{t=0}^{n_h-1} p_{i-1}^t (1-p_{i-1})^{n_h-1-t}
[(n_h-t) \tilde{e}^{(h)}_{n_h-t} \notag \\ 
&\, - (t+1) \tilde{e}^{(h)}_{n_h-1-t}]
\end{align}
where $\tilde{e}^{(h)}_g$ is the $g$th unnormalized information function of~$\mathscr{C}_h$.
\end{proposition}
\medskip
The proof of Proposition~\ref{prop:qhi} is omitted due to space constraints. Note that the proof
follows exactly the same argument used in \cite[Theorem~2]{Ashikhmin:AreaTheorem} to derive the
expression of the EXIT function of a linear block code without idle bits over the binary erasure
channel.

\medskip
\begin{proposition}
Assume MAP decoding is used at each BN. At the $i$th iteration, let $p_{i-1}$ be the average
probability that an edge
carries an erasure message from the SNs to the BNs, before MAP decoding at the burst
node. Let $q_i$ be the average probability that an edge carries an erasure message from the burst
nodes to the SNs, after MAP decoding at the BNs. Then:
\begin{align}\label{eq:q_i(p_i-1)}
q_i = & \frac{1}{\bar{n}}\sum_{h=1}^{n_c} P_h \sum_{t=0}^{n_h-1} p_{i-1}^t
(1-p_{i-1})^{n_h-1-t} [(n_h-t) \tilde{e}^{(h)}_{n_h-t} \notag\\ 
& - (t+1) \tilde{e}^{(h)}_{n_h-1-t}]\, .
\end{align}
\end{proposition}
\medskip
\begin{IEEEproof}
For all $h \in \{1,\dots,n_c\}$, let $\lambda_h$ be the probability that an edge is connected to a
BN of type $h$. We have
\begin{align}\label{eq:lambda_h}
\lambda_h & = \frac{P_h n_h}{\bar{n}} \, .
\end{align}
The proposition is proved by observing that 
\begin{align}\label{eq:q_i}
q_i & = \sum_{h=1}^{n_c} \lambda_h q^{(h)}_i 
\end{align}
and by incorporating \eqref{eq:q^h_i(p_i-1)} and \eqref{eq:lambda_h} into \eqref{eq:q_i}.
\end{IEEEproof}
\medskip

The following is a well-known result from basic density evolution on the erasure channel for
irregular LDPC codes~\cite{studio3:Richardson_design_of_cap_appr}.
\medskip
\begin{proposition}\label{prop:extrinsic_CN_set}
At the $i$th iteration, let $q_i$ be the average probability that an edge carries an erasure message
from the BNs to
the SNs before decoding at the SNs. Let $p_i$ be the average probability that an edge
carries an erasure message from the SNs to the BNs after IC at the SNs. Then:
\begin{align}\label{eq:p_i(q_i)}
p_i = 1-\rho(1-q_i)\, .
\end{align}
\end{proposition}
\medskip

Incorporating \eqref{eq:q_i(p_i-1)} into \eqref{eq:p_i(q_i)} and recalling \eqref{eq:rho(x)}, we
obtain the nonlinear difference equation 
\begin{align}\label{eq:density_evolution}
p_i = 1 - \exp\Big\{ & -\frac{G}{k} \sum_{h=1}^{n_c} P_h \sum_{t=0}^{n_h-1} p_{i-1}^t
(1-p_{i-1})^{n_h-1-t} \notag \\ 
& \times [(n_h-t) \tilde{e}^{(h)}_{n_h-t} - (t+1) \tilde{e}^{(h)}_{n_h-1-t}]\Big\}
\end{align}
which expresses the evolution of the average probability that an edge carries an erasure message at
the $i$th decoding iteration. The initial value of \eqref{eq:density_evolution} shall be set to
$p_0=1$. The asymptotic threshold $G^*$ of the system is defined as
\begin{align*}
G^* := \sup \{G\geq 0 : p_i\rightarrow 0 \textrm{ as }
i\rightarrow\infty,\,\,p_0=1\} \, .
\end{align*}
The threshold $G^*$ is the supremum $G$ such that, in the asymptotic setting
$M\rightarrow\infty$, the normalized throughput $S$ fulfills $S=G$. For all
values of $G<G^*$ the offered traffic turns into useful throughput and therefore $G^*$ is the
asymptotic peak throughput.

Using standard bifurcation theory, the threshold $G^*$ is equal to the smallest $G>0$ such
that, for some $0\leq x < 1$, $(x,G)$ is a solution to the system of simultaneous equations
\begin{align}
f(x,G) & = x \label{eq:system_threshold_eq1} \\
\frac{\partial f(x,G)}{\partial x} & = 1 \label{eq:system_threshold_eq2}
\end{align}
where 
\begin{align}\label{eq:f_definition}
f(x,G) := 1 - \exp\Big\{ & -\frac{G}{k} \sum_{h=1}^{n_c} P_h
\sum_{t=0}^{n_h-1} a^{(h)}_t \notag \\ & \times x^t (1-x)^{n_h-1-t} \Big\}
\end{align}
and $a^{(h)}_t := (n_h-t) \tilde{e}^{(h)}_{n_h-t} - (t+1)
\tilde{e}^{(h)}_{n_h-1-t}\,$.

\begin{table*}[!t]
\caption{Optimized probability distribution {\boldmath $P$} for IRSA schemes with rates
$1/3$, $2/5$ and $R=1/2$ and optimized probability distribution {\boldmath $Q$} for
CSA schemes with $k=2$ and rates $1/3$, $2/5$, $1/2$ and $3/5$ under the random code
hypothesis.}\label{table:ensembles}
\begin{center}
\begin{tabular}{lcccccc|cc}
\hline\hline
\multicolumn{7}{c|}{\emph{IRSA}} & $G^*$ & $G^*_{\mathsf{sb}}$\\
\hline
        & $(2,1)$    & $(3,1)$    & $(6,1)$ &            &     &
 & \\
$R=1/3$ & $0.554016$ & $0.261312$ & $0.184672$ &       &    
&  & $0.8792$ & $0.9025$\\
$R=2/5$ & $0.622412$ & $0.255176$ & $0.122412$ &     &    
&  & $0.7825$ & $0.8033$\\
$R=1/2$ & $1.000000$ &            &                      &     &    
&  &
$0.5000$ & $0.5000$\\
\hline\hline
\multicolumn{7}{c|}{\emph{CSA $k=2$}} & $\bar{G}^*$ & $\bar{G}^*_{\mathsf{sb}}$\\
\hline
        & $(3,2)$    & $(4,2)$    & $(5,2)$ &  $(8,2)$ & $(9,2)$   & $(12,2)$ \\
$R=1/3$ & $0.088459$ & $0.544180$ & $0.121490$ & $\phantom{0.000000}$  &
$\phantom{0.000000}$ &
$0.245871$ & $0.8678$ & $0.9427$\\
$R=2/5$ & $0.153057$ & $0.485086$ & $0.135499$ &
$0.114235$ & $0.112124$ & $\phantom{0.000000}$
& $0.7965$ & $0.8391$\\
$R=1/2$ &  & $1.000000$ &  &    &       &    &
$0.6556$ & $0.7500$\\
$R=3/5$ & $0.666667$ & $0.333333$ &         &    &       &    & $0.4091$ & $0.4091$\\
\hline\hline
\vspace{0.1mm}
\end{tabular}
\end{center}
\end{table*}

\subsection{Stability}
Difference equations such as \eqref{eq:density_evolution} are often used to model discrete dynamical
systems. These systems are typically analyzed as regard to the stability of their fixed (or
steady-state equilibrium) points. A fixed point $\hat{x}$ of $x_{\ell}=f(x_{\ell-1})$ is known to be
locally stable if there exists $\epsilon>0$ such that $\lim_{\ell\rightarrow\infty}x_{\ell}=\hat{x}$
for all $x_0$ such that $|x_0 - \hat{x}|<\epsilon$. The following well-known result establishes a
necessary and sufficient condition for local stability of a fixed point.
\medskip
\begin{lemma}\label{lemma:stability}
A fixed point $\hat{x}$ of a discrete dynamical system $x_{\ell}=f(x_{\ell-1})$,
where $f:\mathbb{R} \mapsto \mathbb{R}$ is a differentiable and single-valued function, is locally
stable if and only if $| f'(\hat{x}) | < 1$.
\end{lemma}

\medskip
It is readily shown that $p=0$ is a fixed point of \eqref{eq:density_evolution}, corresponding to
successful IC. Therefore we may
apply Lemma \ref{lemma:stability} to study its stability. We obtain the following result.
\medskip
\begin{proposition}[Stability condition]\label{proposition:stability}
For $h \in \{1,\dots,n_c\}$, let $\mathscr{C}_h$ be a $(n_h,k,d^{(h)}_{\min})$ linear block code
employed by each user with probability $P_h$ at each transmission and $A^{(h)}_w$ be the number of
weight-$w$ codewords of $\mathscr{C}_h$. Moreover, let $\underline{d}=\min_h \{d^{(h)}_{\min}\}$
and $\mathcal{D}=\{h:d^{(h)}_{\min}=\underline{d}\}$. If $\underline{d}=2$, then the fixed point
$p=0$ of \eqref{eq:density_evolution} is locally stable if and only if
\begin{align}\label{eq:stability_p}
G < \frac{k}{2 \bar{A}_2}
\end{align}
where $\bar{A}_2=\sum_{h\in\mathcal{D}}P_h A^{(h)}_2$ is the average number of weight-$2$ codewords.
Else, if $\underline{d}\geq 3$, then the fixed point $p=0$ of \eqref{eq:density_evolution} is
stable for any value of $G$. 
\end{proposition}
\begin{IEEEproof}
Let us define again $f(x,G)$ as in \eqref{eq:f_definition} and let
us denote by $\mathcal{S}^{(h)}_g$ the generic $(k \times g)$ matrix obtained by selecting $g$
columns in (any representation of) the generator matrix of $\mathscr{C}_h$, irrespective of the
order of the $g$ columns, and by $\sum_{\mathcal{S}^{(h)}_g}$ the summation over all ${n \choose g}$
matrices $\mathcal{S}^{(h)}_g$. We have:
\begin{align}\label{eq:f_prime_(0)_1}
f'(0) & =\frac{G}{k} \sum_{h=1}^{n_c} P_h a^{(h)}_1 \notag \\
\, & = \frac{2G}{k}\sum_{h=1}^{n_c}
P_h\left[\frac{(n_h-1)\tilde{e}^{(h)}_{n-1}}{2}-\tilde{e}^{(h)}_{n-2}\right] \notag \\
\, & \stackrel{\textrm{(a)}}{=} \frac{2G}{k}\sum_{h=1}^{n_c} P_h\left[k {n_h \choose
n_h-2}-\tilde{e}^{(h)}_{n-2}\right] \notag \\
\, & = \frac{2G}{k}\sum_{h=1}^{n_c}
P_h\sum_{\mathcal{S}^{(h)}_{n-2}}(k-\mathrm{rank}(\mathcal{S}^{(h)}_{n-2})) \notag \\
\, & \stackrel{\textrm{(b)}}{=} \left\{\begin{array}{lll}
\frac{2G}{k}\bar{A}_2 & \textrm{ if } & \underline{d}=2 \\ 0 &
\textrm{ if } & \underline{d}\geq3 \end{array} \right.
\end{align}
In the previous equation list, both (a) and (b) rely on the hypothesis that $d_{\min}\geq 2$ and on
\cite[Proposition 2]{paolini09:stability}.
\end{IEEEproof}
\medskip
The stability condition is a necessary, but in general not sufficient condition for successful
decoding. Note also that the stability condition implies
\begin{align}\label{eq:stability_bound}
G^*\leq\frac{k}{2\bar{A}_2}\, . 
\end{align}
that will be referred to as the stability upper bound, denoted by $G^*_{\mathsf{sb}}$. Note that in
the IRSA case ($k=1$) we have $\bar{A}_2=P_2$, where $P_2$ is the probability to select the
length-$2$ repetition code, and therefore for IRSA we obtain %
$
G^*\leq \frac{1}{2 P_2}
$. %

In the case where $\underline{d}=2$, \eqref{eq:stability_bound} may be achieved with equality.
Indeed, this is the case when $n_c=1$ and the binary linear block code $\mathscr{C}$ employed by
all users is a SPC code.
\medskip
\begin{proposition}\label{prop:SPC}
Let $n_c=1$ and the linear block code $\mathscr{C}$ employed by all users be a $(k+1,k)$ SPC
code. Then
\begin{align}
G^* = \frac{1}{k+1}
\end{align}
and \eqref{eq:stability_bound} is achieved with equality.
\end{proposition}

The proof is easily obtained by recasting
\eqref{eq:system_threshold_eq1} and \eqref{eq:system_threshold_eq2} for the special case of
SPC codes and by showing that $(x,G)=(0,\frac{1}{n})$ is a solution to the system
and that no $G<\frac{1}{n}$ exists such that $(x,G)$ is a solution to the system for any
$0\leq x<1$.

\section{CSA from Random Linear Block Codes}\label{sec:random_hypothesis}
So far the generic user has been assumed to encode, at each transmission, its $k$ information
segments via an $(n_h,k,d_{\min}^{(h)})$ binary linear block code picked randomly with
p.m.f. {\boldmath $P$}$=[P_h]_{h=1}^{n_c}$ from an ensemble of $n_c$ candidate codes. In this
section, we consider a slightly different situation. Specifically, we assume that, at each
transmission, the generic user picks randomly a codeword length $n_s>k$ from the ensemble
$\{n_1,\dots,n_{s_{\max}}\}$ with p.m.f. {\boldmath $Q$}$=[Q_{n_s}]_{s=1}^{s_{\max}}$ and
encodes his $k$
segments through a binary $(k\times n_s)$ generator matrix generated uniformly at random
from the set of all rank-$k$ $(k\times n_s)$ binary matrices representing $(n_s,k)$ linear block
codes without idle bits and with minimum distance at least $2$. We are interested in calculating
the expected threshold $\bar{G}^*$ for this scheme. The advantage of a \emph{random code hypothesis}
is to allow to release the analysis from considering a specific set of $n_c$ codes. 

With respect to the previous case, the expression \eqref{eq:Psi_infty} of $\Psi(x)$ and the
expression \eqref{eq:rho(x)} of $\rho(x)$ remain unchanged, provided the definition of $\bar{n}$ is
updated as $\bar{n}=\sum_{s=1}^{s_{\max}} Q_{n_s} n_s$. Analogously, \eqref{eq:p_i(q_i)} is not
affected by
the random code hypothesis. On the other hand, we now update \eqref{eq:q_i(p_i-1)} by replacing
$q_i$ with its average value $\bar{q}_i$. Denoting by $\mathsf{G}_{(n_s,k)}$ the ensemble of all
rank-$k$ $(k \times n_s)$ binary matrices representing linear block codes without idle bits and
with minimum distance at least $2$, and by $\mathbb{E}_{\mathsf{G}_{(n_s,k)}}$ the expectation
operator over $\mathsf{G}_{(n_s,k)}$, we have
\begin{align*}
\bar{q}_i & = \frac{1}{\bar{n}}\sum_{s=1}^{s_{\max}} Q_{n_s} \sum_{t=0}^{n_s-1} p_{i-1}^t
(1-p_{i-1})^{n_s-1-t} \notag\\
& \times[(n_s-t)\mathbb{E}_{\mathsf{G}_{(n_s,k)}}(\tilde{e}_{n_s-t})-(t+1)
\mathbb{E}_{\mathsf{G}_{(n_s,k)}}(\tilde{e}_{n_s-1-t})]
\end{align*}
where again $\bar{n}=\sum_{s=1}^{s_{\max}} Q_{n_s} n_s$. The expectation
$\mathbb{E}_{\mathsf{G}_{(n_s,k)}}(\tilde{e}_g)$ may be calculated using the following result
developed in \cite{paolini10:random}, where a recursive technique to calculate the functions
$J(k,n,k)$ and $K(k,n,g,u,k)$ is also available.
\medskip
\begin{proposition}
For given positive integers $n$, $k<n$, and $g\leq n$,
$\mathbb{E}_{\mathsf{G}_{(n,k)}}(\tilde{e}_g)$ is given by 
\begin{align}
\mathbb{E}_{\mathsf{G}_{(n,k)}}(\tilde{e}_g) = {n \choose g} \sum_{u=1}^{\min\{k,g\}} u\,
\frac{K(k,n,g,u,k)}{J(k,n,k)}
\end{align}
where $J(k,n,k)$ is the number of rank-$k$ $(k \times n)$ binary matrices without zero columns and
without independent columns, and where $K(k,n,g,u,k)$ is the number of rank-$k$ $(k \times n)$
binary matrices without zero columns, without independent columns and such that the first $g$
columns have rank $u$.\footnote{In this context, a column is said to be independent when deleting
the column from the matrix does not affect the rank of the matrix.}
\end{proposition}

\medskip
The average threshold $\bar{G}^*$ may be calculated by properly updating the simultaneous equations
\eqref{eq:system_threshold_eq1} and \eqref{eq:system_threshold_eq2}. Specifically, defining the
function $\bar{f}(x)$ as
\begin{align}\label{eq:f_avg_definition}
\bar{f}(x,G) := 1 - \exp\Big\{ & -\frac{G}{k} \sum_{s=1}^{s_{\max}} Q_{n_s} \sum_{t=0}^{n_s-1}
\mathbb{E}_{\mathsf{G}_{n_s,k}}(a_t)
\notag \\ 
& \times x^t (1-x)^{n_s-1-t} \Big\}
\end{align}
where
$\mathbb{E}_{\mathsf{G}_{n_s,k}}(a_t)=[(n_s-t)\mathbb{E}_{\mathsf{G}_{(n_s,k)}}(\tilde{e}_{
n_s-t} )-(t+1)$ $\mathbb{E}_{\mathsf{G}_{(n_s,k)}}(\tilde{e}_{n_s-1-t})]$, $\bar{G}^*$ is equal to
the smallest $G>0$ such that, for some $0\leq x<1$, $(x,G)$ is a solution of
\eqref{eq:system_threshold_eq1} and \eqref{eq:system_threshold_eq2}, where now $f(x,G)$ is replaced
by $\bar{f}(x,G)$.

Using a proof technique analogous to that of Proposition~\ref{proposition:stability}, it is easy to
show that the stability bound is still given by \eqref{eq:stability_bound}, where now
$\bar{A}_2=\sum_{s=1}^{s_{\max}} Q_{n_s} \bar{A}_2^{(n_s,k)}$ and 
$$\bar{A}_2^{(n_s,k)}\!=\!{n_s \choose 2}\!\! \left(k-\!\!\sum_{u=1}^{\min\{k,n_s-2\}}\!\!u\,
\frac{K(k,n_s,n_s-2,u,k)}{J(k,n_s,k)}\right)$$
is the expected number of weight-$2$ codewords of an $(n_s,k)$ linear block code picked uniformly at
random in the ensemble of all $(n_s,k)$ linear block codes without idle bits and with minimum
distance at least $2$.

\section{Numerical Threshold Optimization and Comparison with
IRSA}\label{sec:numerical_results}

The analysis tool developed in Section~\ref{sec:density_evolution} allows to calculate the
threshold for a given choice of the $n_c$ linear block codes $\mathscr{C}_h$,
$h\in\{1,\dots,n_c\}$, and of the p.m.f. {\boldmath $P$}. Analogously, the tool developed in
Section~\ref{sec:random_hypothesis} allows to evaluate the threshold of a CSA scheme under the
random code hypothesis, for a given choice of the $s_{\max}$ lengths $n_s$,
$n\in\{1,\dots,s_{\max}\}$, and of the p.m.f. {\boldmath $Q$}. These tools can be exploited to
derive optimal (in the sense of
maximizing the threshold $G^*$) probability distributions {\boldmath $P$} and {\boldmath $Q$} in the
two cases. 

Some optimized distribution profiles are shown in Table~\ref{table:ensembles}. Among the
several possible algorithms available to find the global maximum of a nonlinear function,
differential evolution \cite{storn05:DifferentialBook} has been used. In the upper part of the
table, optimized probability distributions {\boldmath $P$} are reported for an IRSA scheme
with rates $1/2$, $2/5$ and $1/3$, while in the lower part optimized probabilities distributions
{\boldmath $Q$} are illustrated for a CSA scheme with $k=2$ and with the same rates,
with the inclusion of $R=3/5$, under the random code hypothesis. All distributions have been
optimized under the constraint that the smallest local rate allowed for each user is $1/6$.
For each distribution, the corresponding threshold $G^*$ and stability bound (right-hand side of
\eqref{eq:stability_bound}) are shown. Note that in the CSA case the threshold values are
average values: Specific choices of the codes $\mathscr{C}_h$ may lead to thresholds $G^*$ larger
than $\bar{G}^*$.

From Table~\ref{table:ensembles} we see that CSA is capable to achieve better
thresholds than IRSA for $R=1/2$ and $R=2/5$, while for the lowest rate $R=1/3$ IRSA
exhibits a better threshold. Accordingly, the IRSA scheme
seems to be preferable in the case of low rates $R$ (i.e., for higher values of the excess power
$\Delta P$), while CSA is more interesting for higher
values of $R$ (i.e., when a higher power efficiency is required). Note also that
CSA allows to achieve values of the overall rate $R>1/2$, whereas only low rates $R\leq1/2$
can be obtained from IRSA, unless some users transmit their burst in the MAC frame
with no repetition. (In this latter case, however, no successful iterative IC can be
guaranteed, so that we always have \mbox{$G^*=0$.}) For example, in Table~\ref{table:ensembles}
an optimized CSA
distribution of rate $R=3/5$ is reported. This distribution has no IRSA counterpart.

\begin{figure}[!t]
\psfrag{G}[t]{\small{$G$}} \psfrag{S}[b]{\small{$S$}} \psfrag{Slotted ALOHA}{\tiny{\textsf{Slotted
ALOHA}}} \psfrag{CSA R12 NCSA1000 k2}{\tiny{\textsf{CSA} $R=1/2$ $k=2$}} \psfrag{CSA
R25 NCSA1000 k2}{\tiny{\textsf{CSA} $R=2/5$
$k=2$}} \psfrag{CSA R35 NCSA 1000 k2}{\tiny{\textsf{CSA} $R=3/5$ $k=2$}} \psfrag{CSA R13 NCSA1000
k2}{\tiny{\textsf{CSA} $R=1/3$ $k=2$}} \psfrag{IRSA R12
NSA500}{\tiny{\textsf{IRSA} $R=1/2$}} \psfrag{IRSA R25 NSA500}{\tiny{\textsf{IRSA} $R=2/5$}}
\psfrag{IRSA R13 NSA500}{\tiny{\textsf{IRSA} $R=1/3$}}
\begin{center}
\includegraphics[width=0.95\columnwidth,draft=false]{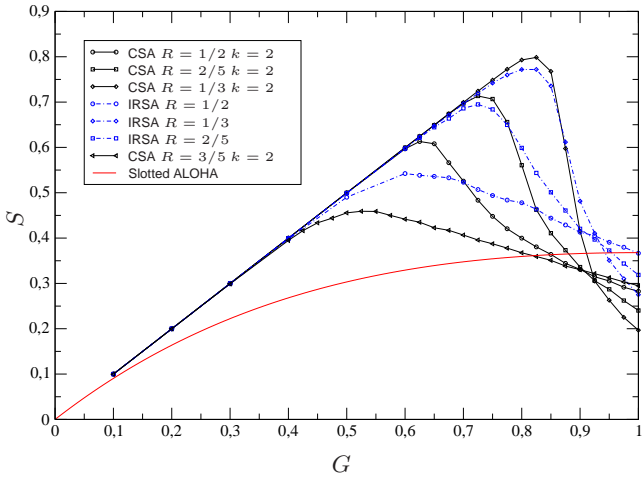}
\end{center}
\caption{Throughput versus the normalized offered traffic $G$ for IRSA
and CSA schemes with p.m.f.'s in Table~\ref{table:ensembles}. The bursts of
each CSA user are split into $k=2$ segments, so that $N_{\rm CSA}=2 N_{\rm SA}$. $N_{\rm
CSA}=1000$, $N_{\rm SA}=500$.}\label{fig:throughput}
\end{figure}

To validate our asymptotic analysis, we performed numerical simulations in the case
of a finite number $M$ of users.\footnote{Here, one shall consider that each segment has to be
encoded via a physical layer error correcting code before transmission on the MAC channel, and
that the
physical layer code for CSA is $k$ times shorter than the corresponding code for
IRSA. Thus, CSA may
require working at slightly higher SNRs than IRSA, especially when short
segments (and then short physical layer codes) are used. This aspect is not
considered in this work.} In Fig.~\ref{fig:throughput}, the throughput
curves of IRSA and CSA schemes with the probability profiles from
Table~\ref{table:ensembles} are depicted as functions of the normalized offered traffic $G$. The
throughput achieved by SA, $S=Ge^{-G}$, is also shown for reference. Note that in our
simulations for the CSA case, we combined the p.m.f.'s {\boldmath $Q$} derived under the
random code hypothesis with a specific choice of the component codes. In particular, we used linear
block codes with the following generator matrices: $\mathbf{G}_{(3,2)}=[110,011]$,
$\mathbf{G}_{(4,2)}=[1100,0111]$, $\mathbf{G}_{(5,2)}=[11100,00111]$,
$\mathbf{G}_{(8,2)}=[111100000,0111111]$, $\mathbf{G}_{(9,2)}=[111110000,011111111]$,
$\mathbf{G}_{(12,2)}=[111111110000,0000011111111]$. To stay fair, we compared CSA ($k=2$) and
IRSA schemes for the same frame duration $T_{\rm F}$, which implies a number of slots $N_{\rm
CSA}$ twice the number of slots $N_{\rm SA}$. Specifically, the simulations are for $N_{\rm
CSA}=1000$ and $N_{\rm SA}=500$. For each value of $G$, $M$ can be obtained from
\eqref{eq:offered_traffic}. We observe a very good match between the asymptotic analysis and
the simulations, the larger peak throughput of CSA than IRSA also for $R=1/3$ being
essentially due to the specific choice of the component codes (recall that $\bar{G}^*$ is an average
value).

\section{Conclusion}\label{sec:conclusion}
Coded slotted ALOHA has been introduced as a new opportunity for high-throughput random access
to the MAC channel. Density evolution equations for CSA have been derived, optimal CSA schemes
designed for several rates and their performance for a finite number of users simulated. The new
scheme is promising when power efficiency is required.



\end{document}